\documentclass{article}
\def\Ref#1{(\ref{#1})}
\usepackage{amsmath}
\usepackage{amssymb}
\usepackage{cite}
\def\d{{\rm d}}
\newcommand{\A}{\operatorname{\cal A}}
\begin{document}
\begin{titlepage}
\noindent{\large\textbf{Dynamical phase transition in the
two-point functions of the autonomous one-dimensional
single-species reaction-diffusion systems}}

\vskip 2 cm

\begin{center}{Amir~Aghamohammadi{\footnote
{mohamadi@azzahra.ac.ir}} \& Mohammad~Khorrami{\footnote
{mamwad@mailaps.org}}} \vskip 5 mm

\textit{     Department of Physics, Alzahra University,
             Tehran 19834, Iran. }

\end{center}

\begin{abstract}
\noindent The evolution of the two-point functions of autonomous
one-dimensional single-species reaction-diffusion systems with
nearest-neighbor interaction and translationally-invariant initial
conditions is investigated. It is shown that the dynamical phase
structure of such systems consists of five phases. As an example,
a one-parameter family is introduced which can be in each of these
phases.
\end{abstract}
\begin{center} {\textbf{PACS numbers:}} 05.40.-a, 02.50.Ga

\end{center}
\end{titlepage}
\section{Introduction}
Reaction-diffusion systems, is a well-studied area. People have
studied reaction-diffusion systems, using analytical
techniques, approximation methods, and simulation. The
approximation methods may be different in different dimensions, as
for example the mean field techniques, working good for high
dimensions, generally do not give correct results for low
dimensional systems. A large fraction of analytical studies,
belong to low-dimensional (specially one-dimensional) systems, as
solving low-dimensional systems should in principle be easier.
\cite{ScR,ADHR,KPWH,HS1,PCG,HOS1,HOS2,AL,AKK,RK,RK2,AKK2,AM1}.

The term exactly-solvable have been used with different meanings.
In \cite{AA,RK3,RK4}, integrability means that the $N$-particle
conditional probabilities' S-matrix is factorized into a product
of 2-particle S-matrices. In
\cite{BDb,BDb1,BDb2,BDb3,Mb,HH,AKA,KAA,MB,AAK}, solvability means
closedness of the evolution equation of the empty intervals (or
their generalization.

In \cite{GS}, a ten-parameter family of reaction-diffusion
processes was introduced for which the evolution equation of
$n$-point functions contains only $n$- or less- point functions.
We call such systems autonomous. The average particle-number in
each site has been obtained exactly for these models. In
\cite{AAMS,SAK}, this has been generalized to multi-species
systems and more-than-two-site interactions.

Among the important aspects of reaction-diffusion systems, is the
phase structure of the system. The static phase structure concerns
with the time-independent profiles of the system, while the
dynamical phase structure concerns with the evolution of the
system, specially its relaxation behavior. In
\cite{MA1,AM2,MAM,MA2}, the phase structure of some classes of
single- or multiple-species reaction-diffusion systems have been
investigated. These investigations were bases on the one-point
functions of the systems.

Here we want to study the two-point functions of autonomous
single-species one-dimensional reaction-diffusion systems.
Throughout this study, the initial condition of the system is
taken to be translationally-invariant, so that it remains
translational-invariant during the evolution. The two-point
function for such systems is obtained, and it is shown that it
exhibits a non-trivial dynamical phase structure. In section 2,
the evolution equation of the two-point function is obtained. In
section 3, this equation is solved and the corresponding
energy-spectrum is obtained. In section 4, the parameter space of
the system is analyzed. In section 5, the dynamical phase
structure of the system (the different phase regions in the
parameter space) is investigated. Finally, section 6 is devoted to
a one-parameter example family, which can be in all five phases.
\section{Evolution equations of the one- and two-point functions}
Consider a one-dimensional periodic lattice, every point of which
is empty or contains one particle. Let the lattice have $L+1$
sites. The observables of such a system are the operators
$N_i^\alpha$, where $i$ with $1\leq i\leq L+1$ denotes the site
number, and $\alpha=0,1$ denotes the hole or the particle: $N_i^0$
is the hole (vacancy) number operator at site $i$, and $N_i^1$ is
the particle number operator at site $i$. One has obviously the
constraint
\begin{equation}\label{1}
s_\alpha N^\alpha_i=1,
\end{equation}
where ${\mathbf s}$ is a covector the components of which
($s_\alpha$'s) are all equal to one. The constraint \Ref{1},
simply says that every site is either occupied by one particle or
empty. A representation for these observables is
\begin{equation}\label{2}
N_i^\alpha:=\underbrace{1\otimes\cdots\otimes 1}_{i-1}\otimes
N^\alpha\otimes\underbrace{1\otimes\cdots\otimes 1}_{L+1-i},
\end{equation}
where $N^\alpha$ is a diagonal $2\times 2$ matrix the only nonzero
element of which is the $\alpha$'th diagonal element, and the
operators 1 in the above expression are also $2\times 2$ matrices.
It is seen that the constraint \Ref{1} can be written as
\begin{equation}\label{3}
{\mathbf s}\cdot{\mathbf N}=1,
\end{equation}
where ${\mathbf N}$ is a vector the components of which are
$N^\alpha$'s. The state of the system is characterized by a vector
\begin{equation}\label{4}
{\mathbf P}\in\underbrace{{\mathbb V}\otimes\cdots\otimes{\mathbb
V}}_{L+1},
\end{equation}
where ${\mathbb V}$ is a $2$-dimensional vector space. All the
elements of the vector ${\mathbf P}$ are nonnegative, and
\begin{equation}\label{5}
{\mathbf S}\cdot{\mathbf P}=1.
\end{equation}
Here ${\mathbf S}$ is the tensor-product of $L+1$ covectors
${\mathbf s}$.

As the values of the number operators $N^\alpha_i$ are zero or
one (and hence $N^\alpha_i$'s are idempotent), the most general
observable of such a system is the product of some of these
number operators, or a sum of such terms. Moreover, the
constraint \Ref{1} shows that the two components of
${\mathbf N}_i$ are not independent. so, one can express any
function of ${\mathbf N}_i$ in terms of
\begin{equation}\label{6}
n_i:={\mathbf a}\cdot{\mathbf N}_i,
\end{equation}
where ${\mathbf a}$ is an arbitrary covector not parallel to
${\mathbf s}$. Our aim is to study the evolution of the two-point
functions constructed by $n_i$'s.

The evolution of the state of the system is given by
\begin{equation}\label{7}
\dot{\mathbf P}={\mathcal H}\;{\mathbf P},
\end{equation}
where the Hamiltonian ${\mathcal H}$ is stochastic, by which it is
meant that its nondiagonal elements are nonnegative and
\begin{equation}\label{8}
{\mathbf S}\; {\mathcal H}=0.
\end{equation}
The interaction is nearest-neighbor, if the Hamiltonian is of the
form
\begin{equation}\label{9}
{\mathcal H}=\sum_{i=1}^{L+1}H_{i,i+1},
\end{equation}
where
\begin{equation}\label{10}
H_{i,i+1}:=\underbrace{1\otimes\cdots\otimes 1}_{i-1}\otimes H
\otimes\underbrace{1\otimes\cdots\otimes 1}_{L-i}.
\end{equation}
(It has been assumed that the sites of the system are identical,
that is, the system is translation-invariant. Otherwise $H$ in the
right-hand side of \Ref{10} would depend on $i$.) The two-site
Hamiltonian $H$ is stochastic, that is, its non-diagonal elements
are nonnegative, and the sum of the elements of each of its
columns vanishes:
\begin{equation}\label{11}
({\mathbf s}\otimes{\mathbf s})H=0.
\end{equation}
Here $H$ is a $4\times 4$ matrix (as the system under
consideration has two possible states in each site and the
interactions are nearest neighbor). The non-diagonal elements of
$H$ are nonnegative and equal to the interaction rates; that is,
the element $H^\alpha_\beta$ with $\alpha\ne\beta$ is equal to the
rate of change of the state $\beta$ to the state $\alpha$.
$\alpha$ and $\beta$, each represent the state of two adjacent
sites. For example if $\alpha=01$ and $\beta=10$, then
$H^\alpha_\beta$ is the rate of particle diffusion to the right.

Using
\begin{equation}\label{12}
{\mathbf s}\otimes{\mathbf s}({\mathbf a}\cdot{\mathbf
N})\otimes({\mathbf b}\cdot{\mathbf N})H=a_\alpha\, b_\beta\,
H^{\alpha\beta}{}_{\gamma\delta}{\mathbf s}\otimes{\mathbf
s}N^\gamma\otimes N^\delta,
\end{equation}
where $\mathbf a$ and $\mathbf b$ are arbitrary covectors, one can
write down the evolution equations of the one- and two-point
functions of $n_i$'s. It turns out that in the evolution equation
of the one-point function, there are two-point functions, and in
the evolution-equation of the two-point function, there are three
point functions, unless the reaction rates satisfy the following
constraints~\cite{GS,AAMS,SAK}.
\begin{equation}\label{13}
\sideset{^e}{^\alpha{}_{\gamma\delta}}\A
=\sideset{^e_1}{^\alpha{}_\gamma}\A\,s_\delta+
\sideset{^e_2}{^\alpha{}_\delta}\A\,s_\gamma,
\end{equation}
where
\begin{align}\label{14}
\sideset{^1}{^\alpha{}_{\gamma\delta}}\A:=&s_\beta\,
H^{\alpha\beta}{}_{\gamma\delta}\nonumber\\
\sideset{^2}{^\alpha{}_{\gamma\delta}}\A:=&s_\beta\,
H^{\beta\alpha}{}_{\gamma\delta}.
\end{align}
It can be seen that one can summarize the constraints
\Ref{13} in the compact form
\begin{equation}\label{15}
H\,{\mathbf u}\otimes{\mathbf u}=\lambda\,{\mathbf
u}\otimes{\mathbf u},
\end{equation}
where
\begin{equation}\label{16}
{\mathbf u}:=
  \begin{pmatrix}
    1\\
    -1
  \end{pmatrix},
\end{equation}
and it is obvious that
\begin{equation}\label{17}
{\mathbf s}\cdot{\mathbf u}=0.
\end{equation}

Now, consider a system satisfying the constraints \Ref{13} (or
equivalently \Ref{15}), and take the vector ${\mathbf v}$
satisfying
\begin{align}\label{18}
\left(\sum_{d,e=1}^2\sideset{^d_e}{}\A\right){\mathbf
v}=&0,\nonumber\\
{\mathbf s}\cdot{\mathbf v}&=1,
\end{align}
and the covector ${\mathbf a}$ such that
\begin{equation}\label{19}
{\mathbf a}\cdot{\mathbf u}=1,\qquad {\mathbf a}\cdot{\mathbf
v}=0,
\end{equation}
that is, the basis $\{{\mathbf a},{\mathbf s}\}$ is dual to
$\{{\mathbf u},{\mathbf v}\}$. This choice of $\textbf{a}$
makes the evolution equation of
$\langle\textbf{a}\cdot\textbf{N}\rangle$ homogeneous.
In \cite{AAMS,SAK}, it is shown
that the matrix in the left-hand side of the first equation in
\Ref{18}, has a left eigenvector with the eigenvalue zero. (This
left eigenvector is ${\mathbf s}$.) So it does have a right
eigenvector with the eigenvalue zero as well. That is, there does
exist a vector ${\mathbf v}$ satisfying \Ref{18}. In fact, one can
even find a real vector ${\mathbf v}$ satisfying \Ref{18}. From
now on, ${\mathbf a}$ in \Ref{6} is assumed to satisfy \Ref{19}.

Assume further, that the initial condition is
translational-invariant. This means that the one-point function is
independent of the site, and the two-point function depends on
only the difference of the sites' numbers. It turns out that the
evolution equation for the one-point function is
\begin{equation}\label{20}
\frac{\d f}{\d t}=(\mu+\nu)f,
\end{equation}
where
\begin{equation}\label{21}
f:=\langle n_i\rangle,
\end{equation}
and
\begin{align}\label{22}
\mu=&{\mathbf s}\otimes{\mathbf a}\, H\,{\mathbf u}\otimes{\mathbf
v}+{\mathbf a}\otimes{\mathbf s}\, H\,{\mathbf v}\otimes{\mathbf
u},\nonumber\\
\nu=&{\mathbf s}\otimes{\mathbf a}\, H\,{\mathbf v}\otimes{\mathbf
u}+{\mathbf a}\otimes{\mathbf s}\, H\,{\mathbf u}\otimes{\mathbf
v}.
\end{align}
Also, taking
\begin{equation}\label{23}
F_i:=\langle n_k\, n_{k+i}\rangle,
\end{equation}
one arrives at
\begin{align}\label{24}
\frac{\d F_i}{\d t}&=\mu(F_{i-1}+F_{i+1})+2\nu\, F_i,\qquad
1<i<L\nonumber\\
\frac{\d F_1}{\d t}&=\mu\, F_2+(\nu+\lambda)F_1+\rho\, f+\sigma,
\end{align}
where
\begin{align}\label{25}
\rho:=&{\mathbf a}\otimes{\mathbf a}\, H\,({\mathbf
u}\otimes{\mathbf v}+{\mathbf v}\otimes{\mathbf u}),\nonumber\\
\sigma:=&{\mathbf a}\otimes{\mathbf a}\, H\,{\mathbf
v}\otimes{\mathbf v}.
\end{align}
From the definition \Ref{23}, it is also seen that
\begin{equation}\label{26}
F_{L+1-i}=F_i.
\end{equation}
It is seen that only five parameters enter the evolution equation
of the up-to-two-point functions, and all of these can be
expressed in terms of the matrix elements of
\begin{equation}\label{27}
\bar H:=H+\Pi\, H\,\Pi,
\end{equation}
where $\Pi$ is the permutation matrix. These parameters can be
rewritten as
\begin{align}\label{28}
\mu:=&{\mathbf s}\otimes{\mathbf a}\, {\bar H}\,{\mathbf
u}\otimes{\mathbf v}\nonumber\\
\nu:=&{\mathbf s}\otimes{\mathbf a}\, {\bar H}\,{\mathbf
v}\otimes{\mathbf u}\nonumber\\
\lambda:=&\frac{1}{2}{\mathbf a}\otimes{\mathbf a}\, {\bar
H}\,{\mathbf u}\otimes{\mathbf u}\nonumber\\
\rho:=&{\mathbf a}\otimes{\mathbf a}\, {\bar H}\,{\mathbf
u}\otimes{\mathbf v}\nonumber\\
\sigma:=&\frac{1}{2}{\mathbf a}\otimes{\mathbf a}\, {\bar
H}\,{\mathbf v}\otimes{\mathbf v}
\end{align}
\section{Solution of the evolution equations}
The solution to \Ref{20} (the evolution equation of the one-point
function) is easily seen to be
\begin{equation}\label{29}
f(t)=f(0)\exp[(\mu+\nu)t].
\end{equation}
Putting this in \Ref{24}, the second equation becomes
\begin{equation}\label{30}
\frac{\d F_1}{\d t}=\mu\,
F_2+(\nu+\lambda)F_1+\rho\,f(0)\exp[(\mu+\nu)t]+\sigma.
\end{equation}
This, combined with the first equation of \Ref{24}, and the
constraint \Ref{26}, are sufficient to obtain the two-point
functions from their initial value. To do so, one takes a solution
like
\begin{equation}\label{31}
F_{i}(t)=\sum_E F_{i\,E}(0)\exp(E\, t),
\end{equation}
and puts it in the equations. From the first equation of \Ref{24},
one arrives at
\begin{equation}\label{32}
E\,F_{i\,E}(0)=\mu[F_{i-1\, E}(0)+F_{i+1\, E}(0)]+ 2\nu\, F_{i\,
E}(0),\qquad 1<i<L.
\end{equation}
\Ref{30} becomes
\begin{equation}\label{33}
E\,F_{1\,E}(0)=\mu\, F_{2\, E}(0)+(\nu+\lambda)F_{1\,E}(0)+\rho\,
f(0)\,\delta_{\mu+\nu,E}+\sigma\,\delta_{0,E}.
\end{equation}
To solve \Ref{32}, one takes
\begin{equation}\label{34}
F_{i\,E}(0)=c_E\, z^i+d_E\, z^{'i}.
\end{equation}
Putting this in \Ref{32}, one arrives at
\begin{equation}\label{35}
E=\mu(z+z^{-1})+2\nu.
\end{equation}
The equation for $z'$ is similar, and in fact $z'$ is the inverse
of $z$. Then, using \Ref{26}, one can write \Ref{34} as
\begin{equation}\label{36}
F_{i\,E}(0)=c_E(z^i+z^{L+1-i}).
\end{equation}
Putting this in \Ref{33}, one can obtain the coefficient $c_E$.

For $E=\mu+\nu$, or $E=0$, the equation for $c_E$ is a
nonhomogeneous one, and $c_E$ is obtained. For $E$ different to
the above values, the equation for $c_E$ is a homogeneous one, and
only for certain values of $E$ there exist nonzero solutions for
$c_E$. These values of $E$ are among the eigenvalues of ${\mathcal
H}$, of course. Using \Ref{36}, \Ref{33}, and \Ref{35}, the
condition for $c_E$ being nonzero is seen to be
\begin{equation}\label{37}
\mu[z^{-(L+1)/2}+z^{(L+1)/2}]=(\lambda-\nu)
[z^{-(L-1)/2}+z^{(L-1)/2}].
\end{equation}
Some of the roots of this equation (for $z$) are phases (their
absolute value is one). In the thermodynamic limit ($L$ to
infinity) it is easy to find the nonphase solutions. It is seen
that if a solution has absolute value less than one, then in the
thermodynamic limit,
\begin{equation}\label{38}
z=\frac{\mu}{\lambda-\nu},
\end{equation}
and it is obvious that such roots of \Ref{37} are real. So, in the
thermodynamic limit, the energy values $E$ entering the
translationally-invariant two point function are zero,
$(\mu+\nu)$, the values coming from \Ref{35} with $|z|=1$, and
possibly only one other value coming from \Ref{35} with $z$
satisfying \Ref{38}. The largest nonzero value of $E$, determines
the relaxation time towards the equilibrium.

One point should be noted. In general, the limit of the largest
relaxation time of a finite system, as its size tends to infinity,
may differ from the relaxation time of the infinite system. It can
be shown, however, that it is not the case for our system. In
fact, if one solves the eigenvalue equation for the infinite
system, one has to omit the periodicity condition \Ref{26}, and
use instead a condition that the two-point function does not blow
up in the limit that the distance between the two points tends to
infinity. This means that either $z$ and $z'$ (the inverse of $z$)
are unimodular, or in \Ref{34} there remains only one term, the
term corresponding to the one with modulus less than one. For the
latter case, one again recovers \Ref{38}. So, the spectrum of the
infinite system, is in fact equal to the limit of the spectrum of
the finite system with periodic boundary conditions, in the
infinite-size limit. It is of course true that if in the initial
condition, the coefficients of some eigenvectors vanish, then the
relaxation time may differ from the largest relaxation time. But
this happens independent of the size of the system. Another case
when the relaxation behavior of the infinite system differs from
the limit that of the finite system, is when the spectrum becomes
continuous to zero, that is, in the infinite limit system, there
is no eigenvalue gap between zero and the other part of the
spectrum. In this case, the relaxation behavior of the infinite
system may be a power law, rather than exponential decaying. But
again this is not the case for the present system. To summarize,
in the present system the largest relaxation time of the infinite
system is equal to the limit of of the largest relaxation time of
the finite system.
\section{The parameter space determining the energy-spectrum of
the two-point functions} Consider a one-dimensional single-species
nearest-neighbor-interacting system, for which the evolution
equations of up-to-$n$-point functions are autonomous. (we call
such systems autonomous.) The Hamiltonian $H$ characterizing such
a system (hence satisfying \Ref{15}), contains 10 parameters. As
it was seen from the previous section, of the parameters entering
$H$, only five parameters enter in the evolution equation of the
two-point functions. All of these are expressible in terms of the
symmetrized (with respect to permutation) Hamiltonian $\bar H$. It
is easily seen that as $H$ satisfies \Ref{15}, $\bar H$ satisfies
\Ref{15} as well. So the system characterized by $\bar H$, is
autonomous as well. Such a system contains 6 independent rates. In
fact, one can write $\bar H$ as
\begin{equation}\label{39}
\bar H=\begin{pmatrix}
       -2 r_1-r_2&r_3&r_3&r_5\\
       r_1&-r_3-r_7-r_4&r_7&r_6\\
       r_1&r_7&-r_3-r_7-r_4&r_6\\
       r_2&r_4&r_4&-r_5-2 r_6
       \end{pmatrix},
\end{equation}
with
\begin{equation}\label{40}
r_1+r_2+r_3=r_4+r_5+r_6.
\end{equation}
Of the five parameters entering the evolution equation of the
two-point function, only three parameters determine the
energy-spectrum. These are $\mu$, $\nu$, and $\lambda$:
\begin{align}\label{41}
\mu&=r_7+r_4-r_1-r_2=r_7+r_3-r_5-r_6,\nonumber\\
\nu&=-r_7-r_1-r_2-r_3=-r_7-r_4-r_5-r_6,\nonumber\\
\lambda&=-\frac{r_1+r_3+r_4+r_6}{2}.
\end{align}
From these relations, it is seen that
\begin{align}\label{42}
\nu&\leq -|\mu|\leq 0,\nonumber\\
\nu&\leq\lambda\leq 0.
\end{align}
As the rates are nonnegative, if $\nu=0$, then $\bar H=0$, which
makes the two-point functions constant. Assuming $\nu\ne 0$, one
can scale time and make $\nu=-1$. So, apart from a time-scale,
there are only two parameters determining the energy-spectrum,
$\mu$ and $\lambda$:
\begin{align}\label{43}
|\mu|&\leq 1,\nonumber\\
-1&\leq\lambda\leq 0,\nonumber\\
\nu&=-1.
\end{align}
It can be shown that the whole region of the above is physical.
That is, corresponding to any $\lambda$ and $\mu$ satisfying the
above inequalities, there are autonomous systems yielding the
desired $\lambda$ and $\mu$. To prove this, first consider four
specific systems:
\begin{itemize}
\item $r_1=r_6=1,\quad
r_2=r_3=r_4=r_5=r_7=0,\quad\Rightarrow\quad(\mu,\lambda)=(-1,-1)$.
\item $r_2=r_5=1,\quad
r_1=r_3=r_4=r_6=r_7=0,\quad\Rightarrow\quad (\mu,\lambda)=(-1,0)$.
\item $r_3=r_4=1,\quad
r_1=r_2=r_5=r_6=r_7=0,\quad\Rightarrow\quad (\mu,\lambda)=(1,-1)$.
\item $r_7=1,\quad
r_1=r_2=r_3=r_4=r_5=r_6=0,\quad\Rightarrow\quad
(\mu,\lambda)=(1,0)$.
\end{itemize}
Any point $(\mu,\lambda)$ in the region described by \Ref{43}, can
be written as
\begin{equation}\label{44}
(\mu,\lambda)=c_1(-1,-1)+c_2(-1,0)+c_3(1,-1)+c_4(1,0),
\end{equation}
where $c_a$'s are nonnegative. A system with the rates
\begin{equation}\label{45}
r_i=\sum_{a=1}^4 c_a\,r_{ai},
\end{equation}
where $r_{ai}$ is the rate $r_i$ of the $a$'th system introduced
above, gives the desired $(\mu,\lambda)$.
\section{Dynamic phase transitions in the two-point function}
It was shown in the previous section, that for any nonzero
Hamiltonian $\nu\ne 0$, so that one can normalize $\nu$ to $-1$.
In section 3, it was shown that the energies entering the
two-point function are $0$, $E_1:=\mu-1$, and $\mu(z+z^{-1})-2$,
where in the thermodynamic limit $|z|=1$, or at most one
non-unimodular $z$ exists, the value of which comes from \Ref{38}.
This is provided the absolute-value of the left-hand side of
\Ref{38} is less than one. So, the energies (apart from $0$) are
$E_1$, any number in the interval $I_0:=[-2-2|\mu|,-2+2|\mu|]$,
and possibly
\begin{equation}\label{46}
E_2:=\lambda-1+\frac{\mu^2}{\lambda+1},
\end{equation}
The largest relaxation time of the two-point function is
$-E_{\textrm{max}}^{-1}$, where $E_{\textrm{max}}$ is the largest
nonzero value of the energy spectrum. The relaxation time of the
one-point function is $-E_1^{-1}$. The fact that the energy
spectrum of the one-point function consists of a single value, is
a result of the translational-invariance of the initial state of
the system. Otherwise, there would be many energies for the
one-point function, which could lead to a dynamical
phase-transition in the one-point function \cite{MA1,AM2,MAM,MA2}.
The comparison of the relaxation times of the two-point- and the
one-point-functions, is a comparison of $E_{\textrm{max}}$ and
$E_1$. If the former is larger, the largest relaxation time of the
two-point function is larger than the relaxation time of the
one-point function (the slow phases). If the two are equal, the
relaxation-times are equal (the fast phases). So, the relation of
$E_1$, $E_2$, and $I_0$, determines the relaxation behavior of the
two-point function (its dynamical phase). It is seen that
\begin{align}\label{47}
&I_0<E_1,\qquad \mu>-\frac{1}{3},\nonumber\\
&E_1\in I_0,\qquad \mu<-\frac{1}{3},
\end{align}
where $|\mu|\leq 1$ has also been used. If $E_1>I_0$, then the
relaxation time of the one-point function is equal to the largest
relaxation time of the two-point function. If $E_1\in I_0$, the
the largest relaxation time of the two-point function is larger.

For $E_2$ to be among the energies, the absolute value of the
left-hand side of \Ref{38} should be less than one. So,
\begin{align}\label{48}
\not\exists &E=E_2,\qquad \lambda<|\mu|-1,\nonumber\\
\exists &E=E_2,\qquad \lambda>|\mu|-1.
\end{align}

Finally,
\begin{equation}\label{49}
(\lambda+1)(E_2-E_1)=\lambda^2+\mu^2-\lambda\,\mu+\lambda-\mu=:
f(\mu,\lambda),
\end{equation}
from which (using $\lambda+1$ is nonnegative),
\begin{align}\label{50}
&E_2<E_1,\qquad f(\mu,\lambda)<0,\nonumber\\
&E_2>E_1,\qquad f(\mu,\lambda)>0.
\end{align}
$f=0$ is an ellipse, the interior points of which correspond to
$E_2<E_1$, and the exterior points of which correspond to
$E_2>E_1$.

These three inequalities divide the whole phase space $(|\mu|\leq
1,\,\-1\leq\lambda\leq 0)$ into five phases:
\begin{itemize}
\item[\textbf{I)}] $\mu<-\frac{1}{3},\quad \lambda<|\mu|-1.$

In this phase, $E_1\in I_0$, and $E_2$ is not an energy. This is
the slower phase, and the largest energy is
$E_{\textrm{max}}=-2-2\mu$.

\item[\textbf{II)}] $\mu<-\frac{1}{3},\quad \lambda>|\mu|-1.$

In this phase, $E_1\in I_0$, and $E_2$ is an energy, in fact the
largest one. This is the slowest phase, and the largest energy is
$E_{\textrm{max}}=-1+\lambda+\frac{\mu^2}{\lambda+1}$.

\item[\textbf{III)}] $\mu>-\frac{1}{3},\quad \lambda<|\mu|-1.$

In this phase, $E_1> I_0$, and $E_2$ is not an energy. This is the
fastest phase, and the largest energy is
$E_{\textrm{max}}=-1+\mu$.

\item[\textbf{IV)}] $\mu>-\frac{1}{3},\quad |\mu|-1<\lambda< \frac{\mu-1+\sqrt{(1+3\mu)(1-\mu)}}{2}.$

In this phase, $E_1> I_0$, $E_2$ is an energy, and $E_2<E_1$. This
is the fast phase, and the largest energy is
$E_{\textrm{max}}=-1+\mu$.

\item[\textbf{V)}] $\mu>-\frac{1}{3},\quad \lambda> \frac{\mu-1+\sqrt{(1+3\mu)(1-\mu)}}{2}.$

In this phase, $E_1> I_0$, $E_2$ is an energy, and $E_2>E_1$. This
is the slow phase, and the largest energy is
$E_{\textrm{max}}=-1+\lambda+\frac{\mu^2}{\lambda+1}$.
\end{itemize}
This phase structure is summarized in fig. 1.


As previously mentioned, these phases arise from the fact that the
energy-spectrum of the two-point function consists of a continuous
part, an energy equal to the energy appearing in the one-point
function, and possibly another energy. This shows that the
relaxation of the two-point function is at least as slow as that
of the one-point function, and may be slower, depending on the
relative position of the discrete and continuous parts of the
spectrum. The fast phases (phases \textbf{III} and \textbf{IV}),
are those in them the relaxation time of the one- and two-point
functions are equal, while in the slow phases (phases \textbf{I},
\textbf{II}, and \textbf{V}), the relaxation of the two-point
function is slower than that of the one-point function.

\section{A one-parameter family as an example}
Consider a system with the Hamiltonian
\begin{equation}\label{51}
H=\frac{1}{4}
\begin{pmatrix}
-3+3\omega&\omega&\omega&1-\omega\\
1-\omega&-3\omega&\omega&1-\omega\\
1-\omega&\omega&-3\omega&1-\omega\\
1-\omega&\omega&\omega&-3+3\omega\\
\end{pmatrix}.
\end{equation}
This Hamiltonian describes a system with the following reactions.
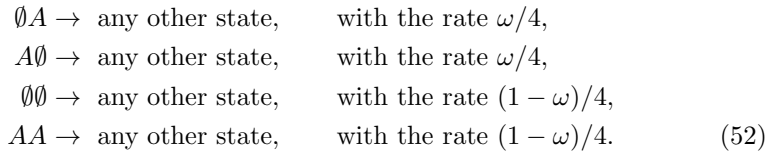
\begin{align}\label{52}
\emptyset A&\to\hbox{ any other state},\qquad\hbox{with the rate
$\omega/4$},\nonumber\\
A\emptyset&\to\hbox{ any other state},\qquad\hbox{with the rate
$\omega/4$},\nonumber\\
\emptyset\emptyset&\to\hbox{ any other state},\qquad\hbox{with the
rate $(1-\omega)/4$},\nonumber\\
AA&\to\hbox{ any other state},\qquad\hbox{with the rate
$(1-\omega)/4$}.
\end{align}
It is seen that for this system,
\begin{equation}\label{53}
\bar H=2H,
\end{equation}
and
\begin{align}\label{54}
\mu&=-1+2\omega,\nonumber\\
\lambda&=-\frac{1}{2}.
\end{align}
For this system, $\rho$ defined through \Ref{28} is equal to zero,
hence there is no term proportional to $e^{E_1\, t}$ in the
right-hand side of \Ref{30}. However, it is seen that adding a
term
\begin{equation}\label{55}
H_1=r
\begin{pmatrix}
-1&0&0&0\\
0&0&0&1\\
0&0&0&1\\
1&0&0&-2\\
\end{pmatrix},
\end{equation}
to the Hamiltonian $H$ in \Ref{51}, changes the values of $\rho$,
$\nu$, $\mu$, and $\lambda$ to
\begin{align}\label{56}
\rho=&r,\nonumber\\
\nu=&-1-2r,\nonumber\\
\mu=&-1-2r+2\omega,\nonumber\\
\lambda=&-\frac{1}{2}-r.
\end{align}
For small values of $r$, one can use $\nu=1$, and $\mu$ and
$\lambda$ as in \Ref{54}. Then, with different values of $\omega$,
this system can exist in all the above five phases:
\begin{align}\label{57}
&\hbox{phase \textbf{I}},\qquad
0\leq\omega<\frac{1}{4},\nonumber\\
&\hbox{phase \textbf{II}},\qquad
\frac{1}{4}<\omega<\frac{1}{3},\nonumber\\
&\hbox{phase \textbf{V}},\qquad
\frac{1}{3}<\omega<\frac{5-\sqrt{5}}{8},\nonumber\\
&\hbox{phase \textbf{IV}},\qquad
\frac{5-\sqrt{5}}{8}<\omega<\frac{3}{4},\nonumber\\
&\hbox{phase \textbf{III}},\qquad \frac{3}{4}<\omega\leq 1.
\end{align}
It is seen that increasing $\omega$, the system undergoes phase
transitions from the phase \textbf{I} to \textbf{II}, then
\textbf{V}, \textbf{IV}, and finally \textbf{III}.


\newpage
\begin{thebibliography}{99}
\bibitem{ScR}  G. M. Sch\"{u}tz; ``Exactly solvable models for many-body
               systems far from equilibrium'' in ``Phase
               transitions and critical phenomena, vol. \textbf{19}'',
               C. Domb \& J. Lebowitz (eds.), (Academic
               Press, London, 2000).
\bibitem{ADHR} F. C. Alcaraz, M. Droz, M. Henkel, \& V. Rittenberg;
               Ann. Phys. \textbf{230} (1994) 250.
\bibitem{KPWH} K. Krebs, M. P. Pfannmuller, B. Wehefritz, \&
               H. Hinrichsen; J. Stat. Phys. \textbf{78}[FS] (1995) 1429.
\bibitem{HS1}  H. Simon; J. Phys. \textbf{A28} (1995) 6585.
\bibitem{PCG}  V. Privman, A. M. R. Cadilhe, \& M. L. Glasser; J. Stat.
               Phys. \textbf{81} (1995) 881.
\bibitem{HOS1} M. Henkel, E. Orlandini, \& G. M. Sch\"utz; J. Phys.
               \textbf{A28} (1995) 6335.
\bibitem{HOS2} M. Henkel, E. Orlandini, \& J. Santos; Ann. of Phys.
               \textbf{259} (1997) 163.
\bibitem{AL}   A. A. Lushnikov; Sov. Phys. JETP \textbf{64} (1986) 811
               [Zh. Eksp. Teor. Fiz. \textbf{91} (1986) 1376].
\bibitem{AKK}  M. Alimohammadi, V. Karimipour, \& M. Khorrami; Phys. Rev.
               \textbf{E57} (1998) 6370.
\bibitem{RK}   F. Roshani \& M. Khorrami; Phys. Rev. \textbf{E60} (1999) 3393.
\bibitem{RK2}  F. Roshani \& M. Khorrami; J. Math. Phys. \textbf{43} (2002) 2627.
\bibitem{AKK2} M. Alimohammadi, V. Karimipour, \& M. Khorrami; J. Stat.
               Phys. \textbf{97} (1999) 373.
\bibitem{AM1}  A. Aghamohammadi \& M. Khorrami; J. Phys. \textbf{A33} (2000) 7843.
\bibitem{AA}   M. Alimohammadi, \& N. Ahmadi; Phys. Rev. \textbf{E62} (2000) 1674.
\bibitem{RK3}  F. Roshani \& M. Khorrami; Phys. Rev. \textbf{E64} (2001) 011101.
\bibitem{RK4}  F. Roshani \& M. Khorrami; Eur. Phys. J. \textbf{B36} (2003) 99.
\bibitem{BDb}  M. A. Burschka, C. R. Doering, \& D. ben-Avraham;  Phys.
               Rev. Lett. \textbf{63} (1989) 700.
\bibitem{BDb1} D. ben-Avraham;  Mod. Phys. Lett. \textbf{B9} (1995)
               895.
\bibitem{BDb2} D. ben-Avraham; in ``Nonequilibrium Statistical
               Mechanics in One Dimension'', V. Privman (ed.), pp 29-50
               (Cambridge University press,1997).
\bibitem{BDb3} D. ben-Avraham; Phys. Rev. Lett. \textbf{81} (1998)
               4756.
\bibitem{Mb}   T. Masser, D. ben-Avraham; Phys. Lett. \textbf{A275} (2000) 382.
\bibitem{HH}   M. Henkel \& H. Hinrichsen; J. Phys. \textbf{A34}, 1561-1568
               (2001).
\bibitem{AKA}  M. Alimohammadi, M. Khorrami, \& A. Aghamohammadi;
               Phys. Rev. \textbf{E64} (2001) 056116.
\bibitem{KAA}  M. Khorrami, A. Aghamohammadi, \& M. Alimohammadi;
               J. Phys. \textbf{A36} (2003) 345.
\bibitem{MB}   M. Mobilia \& P. A. Bares; Phys. Rev. \textbf{E64} (2001) 066123.
\bibitem{AAK}  A.~Aghamohammadi, M.~Alimohammadi, \& M.~Khorrami;
               Eur. Phys. J. \textbf{B31} (2003) 371.
\bibitem{GS}   G. M. Sch\"utz; J. Stat. Phys. \textbf{79}(1995) 243.
\bibitem{AAMS} A. Aghamohammadi, A. H. Fatollahi, M. Khorrami, \& A.
               Shariati; Phys. Rev. \textbf{E62} (2000) 4642.
\bibitem{SAK}  A. Shariati, A. Aghamohammadi, \& M. Khorrami;
               Phys. Rev. \textbf{E64} (2001) 066102.
\bibitem{MA1}  M. Khorrami \& A. Aghamohammadi; Phys. Rev. \textbf{E63} (2001)
               042102.
\bibitem{AM2}  A. Aghamohammadi \& M. Khorrami; J. Phys. \textbf{A34} (2001) 7431.
\bibitem{MAM}  N. Majd, A. Aghamohammadi, \& M. Khorrami;
               Phys. Rev. \textbf{E64} (2001) 046105.
\bibitem{MA2}  M. Khorrami \& A. Aghamohammadi; Phys. Rev. \textbf{E65}
               (2002) 056129.
\end{thebibliography}
\end{document}